\documentclass[12pt]{article}
\pdfoutput=1
\usepackage{jinstpub}
\usepackage{float}
\usepackage{lineno}

\title{Nuclear Recoil Scintillation Linearity of a High Pressure $^4$He Gas Detector}

\author[a,1]{A. Biekert,\note{Corresponding author.}}
\author[b]{S. A. Hertel,}
\author[a]{E. Huebler,}
\author[a]{J. Lin,}
\author[b]{H. D. Pinckney,}
\author[a]{R. K. Romani,}
\author[b]{A. Serafin,}
\author[a]{V. Velan,}
\author[a,c]{and D. N. McKinsey}

\affiliation[a]{University of California Berkeley, Department of Physics, Berkeley, CA 94720, USA}
\affiliation[b]{University of Massachusetts-Amherst, Department of Physics, 1126 Lederle Graduate Research Tower, Amherst, MA 01003-9337 USA}
\affiliation[c]{Lawrence Berkeley National Laboratory, 1 Cyclotron Rd., Berkeley, CA 94720, USA}

\emailAdd{biekerta@berkeley.edu}

\abstract{We investigate scintillation linearity of a commercial high pressure $^4$He gas detector using monoenergetic 2.8 MeV neutrons from a deuterium-deuterium fusion neutron generator. The scintillation response of the detector was measured for a range of recoil energies between 83 keV and 626 keV by tagging neutrons scattering into fixed angles with a far-side organic scintillator detector. Detailed Monte Carlo simulations were compared to experimental data to determine the linearity of the detector response by comparing the scaling of the energy deposits in the simulations to the detector output. In this analysis, a linear scintillation response corresponds to a consistent value for the scaling factor between simulated energy deposits and experimental data for several different scattering angles. We demonstrate that the detector can be used to detect fast neutron interactions down to 83 keV recoil energies and can be used to characterize low-energy neutron sources, one of its potential applications.}

\arxivnumber{1907.03985}

\begin{document}
\maketitle
\flushbottom

\section{Introduction}

    Recent work has shown $^4$He gas can be used as a detection medium for fast neutrons, due to its properties as a scintillator and a barn-scale cross section for the elastic scattering of MeV-scale neutrons \cite{RC2012a, DM2014a}. Fast neutron detectors also require good timing characteristics and gamma-ray rejection capabilities, both of which have led to the widespread use of liquid organic scintillator materials for this application. While organic scintillators are generally engineered for good gamma-ray rejection, they typically have non-linear scintillation response to neutron recoils, particularly for sub-MeV recoil energies \cite{CA2018a}. The scintillation response is an important quantity in reconstructing the energy deposited into the detector on an event-by-event basis, since it can be dependent both on the type of interaction in the detector and the recoil energy. It has previously been shown that the scintillation response of helium gas to neutron recoils is fairly linear down to 380 keV \cite{GM1968a} and more recently to about 240 keV \cite{RK2015a}, making it an ideal candidate for fast neutron detection with good recoil energy reconstruction capabilities. The scintillation response of helium gas has also been investigated for ion recoils \cite{DS2008a}. 
    
    Arktis Radiation Detectors, Ltd.~has produced a commercial product based on a detector volume consisting of $^{4}$He gas instrumented with light collection sensors to detect scintillation from particle recoils in the gas \cite{A2019a}. The first versions of the detectors were instrumented with photomultiplier tubes (PMTs) on both ends of the cylindrical vessel \cite{RC2012a}. Pulse shape discrimination based on the prompt light fraction can distinguish between nuclear recoils from particles such as neutrons and electron recoils, which are background events from environmental gamma-rays and charged particles \cite{RK2015b, RJ2015a}. More importantly, the detector is naturally less sensitive to gamma-rays because the recoiling electron will likely travel far enough to deposit some of its energy into the detector wall instead of the scintillating gas \cite{RC2012a}. Thus, the main discrimination strategy in the modern iteration of the detector is simply to compare the size of the pulse and discard events below a threshold as electron recoil backgrounds \cite{RJ2015a, M2016a}. For neutron recoils, on the other hand, the detector has been used to extract spectral information about neutron sources \cite{RK2015a, JL2014a, TZ2017a}. 
    
    The newer iteration of the Arktis S-670 fast neutron detector is instrumented with silicon photomultipliers (SiPMs) \cite{MC2013b, CB2018a}. The active one liter volume consists 180 bar of $^{4}$He gas in an approximately 1 meter long stainless steel tube with 4 millimeter thick walls. There are three equally-sized, optically isolated segments, each containing four channels corresponding to a SiPM pair for a total of 24 SiPMs in the detector. Each channel can be read out through an analog board with amplification and shaping electronics provided by Arktis \cite{M2015a}. 
    
    In dark matter direct detection experiments, calibration sources which produce low-energy neutron recoils are an active area of research \cite{JC2013a}. Successful neutron recoil calibrations require an understanding of the source spectrum and activity. In this study, we use the SiPM-instrumented Arktis detector to measure the linearity of the detector response to low-energy nuclear recoils, to investigate whether it could be applied to the characterization of neutron sources such as those based on the $^{88}$Y/Be photoneutron reaction \cite{JC2013a}.

\section{Experimental Setup}
    
    \subsection{Neutron Scattering}
    \label{ssec:neutron_scattering}

\begin{figure}[h]
        \centering\includegraphics[width=5.in]{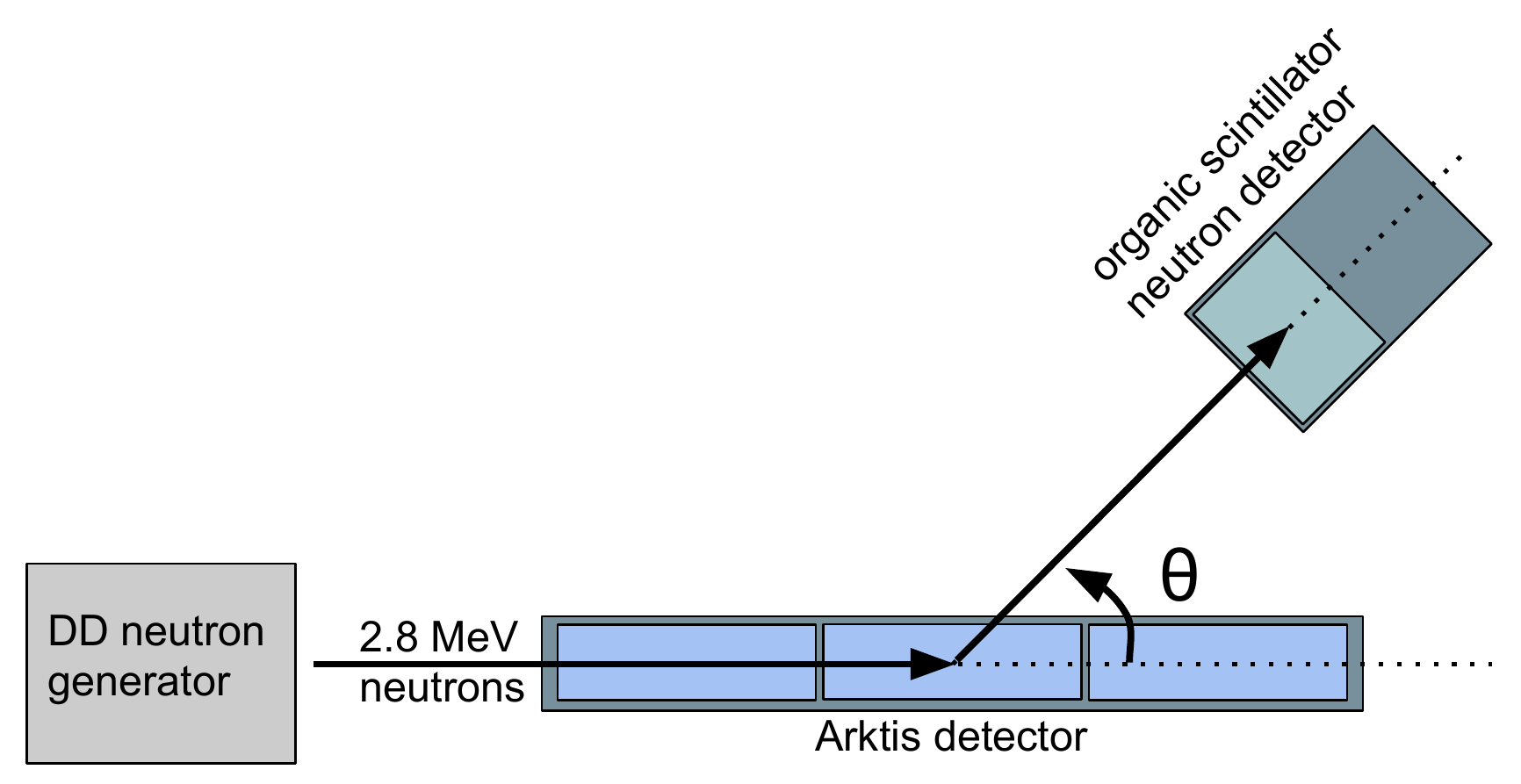}
        \caption{The scattering geometry for measuring the nuclear recoil scintillation linearity of the high pressure helium gas in the Arktis detector. The detectors were positioned about 1 m apart for each configuration; this figure is not drawn to scale.}
        \label{fig:geometry}
    \end{figure}

    Neutrons which elastically scatter from helium atoms deposit energy according to the equation
    \begin{equation}
        E_r = \frac{2 m_n E_n}{(m_n + m_\mathrm{He})^2}\left[m_n \sin^2(\theta) + m_\mathrm{He} - \cos(\theta)\sqrt{m^2_\mathrm{He} - m_n^2\sin^2(\theta)}\right],
        \label{eq:energy_angle}
    \end{equation}
    where $E_r$ is recoil energy of the helium atom, $m_n$ is the mass of the neutron, $E_n$ is the initial energy of the neutron, $m_\mathrm{He}$ is the mass of the helium atom, and $\theta$ is the scattering angle of the neutron in the lab frame relative to its initial direction. We performed a neutron scattering experiment with the Arktis detector and a  St.~Gobain BC-501A organic liquid scintillator \cite{SG2019a} detector with PMT readout and a 5 inch diameter and 5 inch height active volume to detect neutrons produced by an MP320 deuterium-deuterium (DD) neutron generator made by Thermo Fisher Scientific \cite{TF2019a}, which produces 2.8 MeV neutrons in the forward direction \cite{HL1973a}. By tagging neutrons which scatter from the Arktis detector into the organic scintillator detector, positioned at a specific angle as in Figure~\ref{fig:geometry}, we fix the energy deposit into the Arktis detector assuming the neutron scatters only once.
    
    We studied five scattering angles ranging from 20-60 degrees, corresponding to 83-626 keV in mean recoil energy as shown in Table \ref{tab:energy_ranges}. Monte Carlo (MC) simulations were performed with the \textsc{Geant4} toolkit \cite{SA2003a} for the detector geometries shown in Figure \ref{fig:geometry}. In the simulation, the DD generator output was approximated as 2.8 MeV neutrons distributed isotropically from the source plane. The DD generator neutrons in reality had a small dependence in energy and flux with the angle they left the target plane \cite{HL1973a}, but these variations were negligible due to the small solid angle subtended by the Arktis detector volume when placed 1 m away from the DD source plane as in our experiment. Simulated events that deposited energy in both the middle cell of the Arktis detector and the organic scintillator had their recoil energies and deposit times recorded. The expected time of flight between the Arktis and organic scintillator detectors in our configurations was between 44 and 49 ns, depending on the energy of the scattered neutron. Neutrons which scattered more than once in the Arktis detector or other materials, and therefore did not deposit the energy given by Equation~\ref{eq:energy_angle}, contributed about 10\% of events in the single scattering signal region in the simulation. 
    
\begin{figure}[h]
        \centering\includegraphics[width=4.in]{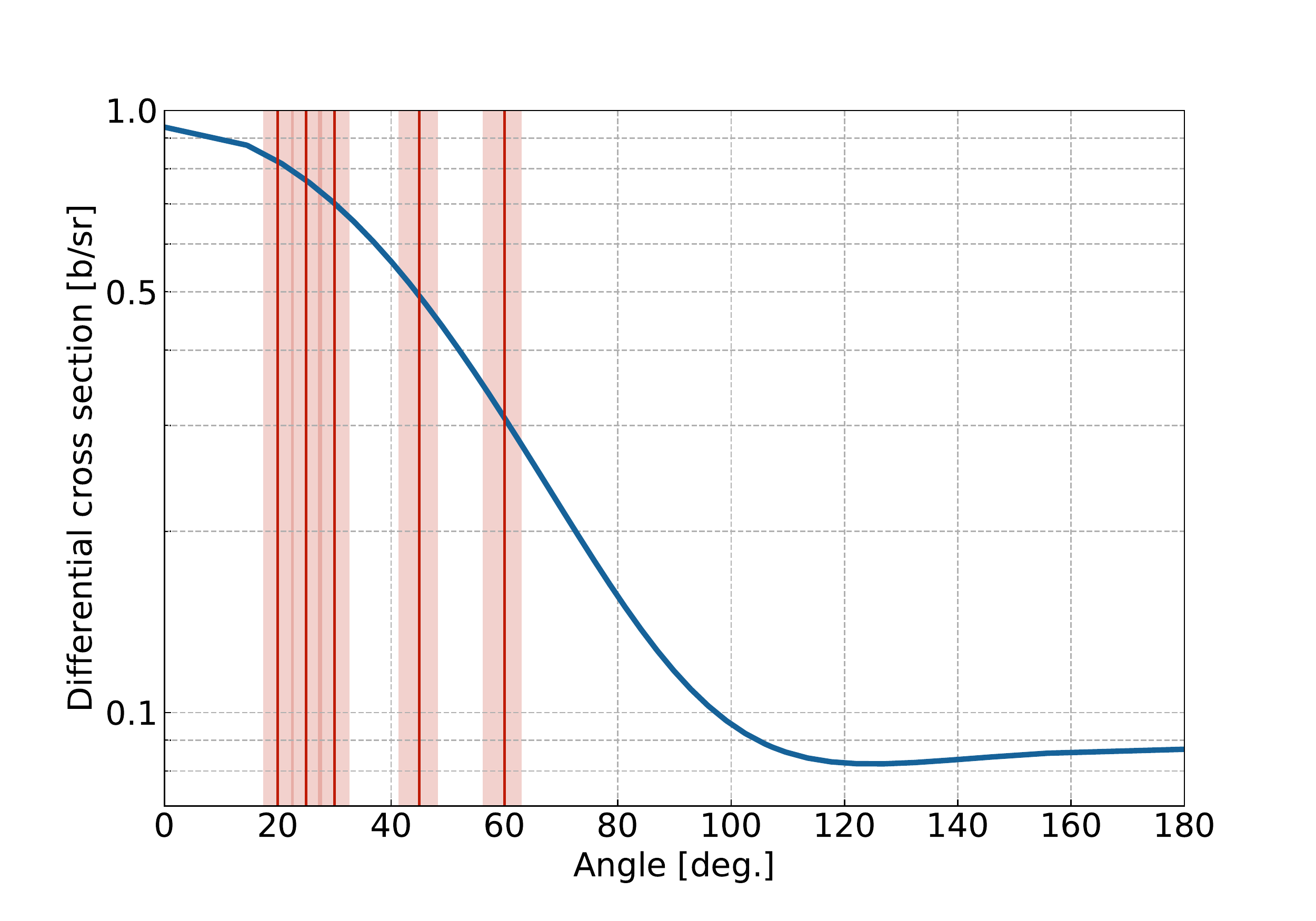}
        \caption{Neutron elastic scattering differential cross section for $^4$He from the ENDF/B-VII library \cite{MC2006a} used in Geant4 simulations. Vertical red lines represent the angles at which data was taken, with shaded bounds corresponding to the $1\sigma$ angular spread of the experimental geometry as discussed in the text.}
        \label{fig:cross_section}
    \end{figure}

    Since the detectors have finite physical size, there is a spread in scattering angles and therefore recoil energies represented in each geometric configuration, even when only considering single scatter events. To estimate the size of this effect, we selected single scattering events from the MC simulations for each experimental geometry and fit a Gaussian to their spread in energy. We report the $1\sigma$ spread in energy and scattering angle in Table \ref{tab:energy_ranges}. We also show the sampled regions of the differential cross section for these five scattering angles in Figure \ref{fig:cross_section}. Since the elongated cell geometry contributes to smearing in the scattering angle, and each cell of the Arktis detector has slightly different optical properties \cite{CB2018a}, we restricted ourselves to using the middle cell of the Arktis detector.
    
    \begin{table}[t]
        \centering
        \caption{The measured scattering angles and the corresponding energy of single-scattering neutrons given by Equation~\ref{eq:energy_angle}, as well as the 1$\sigma$ spread in angles from the finite sizes of detector elements and the corresponding 1$\sigma$ spread in energies.}
        \begin{tabular}{|cc
                        @{\hspace*{15mm}}cc|}
            \hline
            $\theta$ & $\sigma_\theta$ & $E_r$ & $\sigma_{E_r}$\\
            $[\mathrm{deg.}]$ & $[\mathrm{deg.}]$ & $[\mathrm{keV}]$ & $[\mathrm{keV}]$\\
            \hline\hline
            20 & 3 & 83 & 22 \\ 
            25 & 3 & 129 & 27 \\
            30 & 3 & 182 & 32 \\
            45 & 4 & 384 & 52 \\
            60 & 4 & 626 & 58 \\
            \hline
        \end{tabular}
        \label{tab:energy_ranges}
    \end{table}
    
    \begin{figure}[h]
        \centering\includegraphics[width=\linewidth]{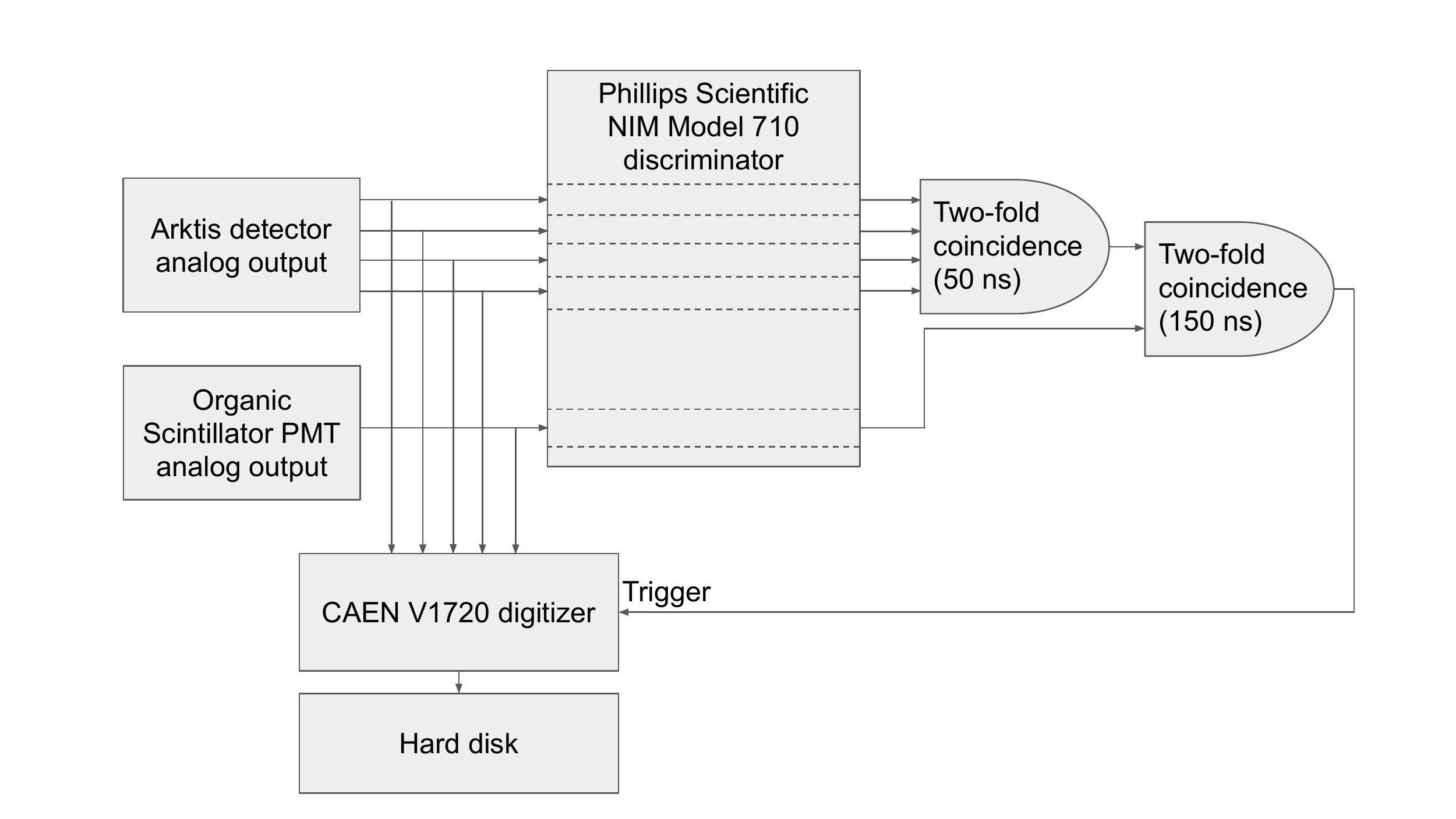}
        \caption{The information flow and trigger logic of the neutron scattering experiment described in section~\ref{ssec:neutron_scattering}. Two-fold coincidence of at least
        two of the four Arktis channels within 50 ns was imposed to increase the probability of triggering on a neutron scatter and not a gamma event or noise. Coincidence between the Artkis detector and the organic scintillator pulses was relaxed to 150 ns to enable a time of flight cut in the analysis step.}
        \label{fig:DAQ_Cartoon}
    \end{figure}
    
\begin{figure}[h]
        \centering\includegraphics[width=5.in]{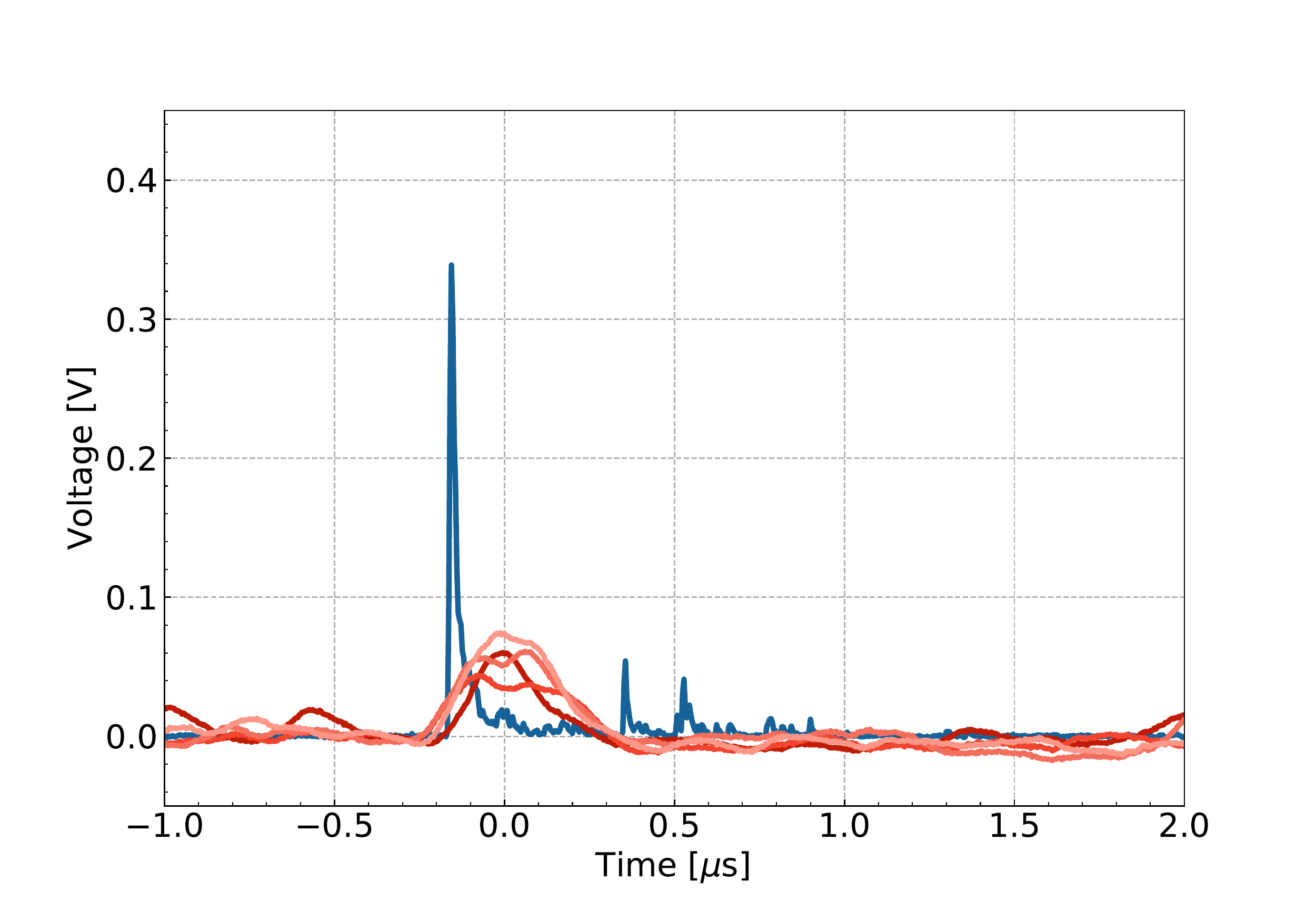}
        \caption{A sample event with pulses from the four channels of the Arktis detector in shades of red and the amplified PMT output of the organic scintillator detector in blue. Negative time corresponds to samples preceding the trigger. The y-axis is the relative voltage recorded by the digitizer after baseline subtraction has been performed on the five detector channels.}
        \label{fig:event}
    \end{figure}

    Pulses were read out of the Arktis and organic scintillator detectors into a NIM module trigger system. A Phillips Scientific NIM Model 710 discriminator was used to determine when two or more of the four channels in the middle segment of the Arktis detector had pulses cross the threshold value within a 50 ns window. This two-fold coincidence requirement eliminates many electron recoil backgrounds, which are less likely to produce pulses in two or more channels at once \cite{CB2018a}. The event was recorded if the Arktis pulse coincidence, the first two-fold coincidence requirement depicted in Figure~\ref{fig:DAQ_Cartoon}, occurred within 150 ns of a pulse in the organic scintillator module, the second two-fold coincidence requirement in Figure~\ref{fig:DAQ_Cartoon}. An extended coincidence window was selected to allow for a time of flight cut in the analysis of the data. When triggered, all detector channels were digitized and recorded by a CAEN V1720 digitizer with a 4 ns sampling rate. A typical neutron-like event with pulses from both detectors is shown in Figure \ref{fig:event}.
    
    \subsection{Trigger Efficiency Measurement}
    
    \begin{figure}[h]
        \centering\includegraphics[width=5.in]{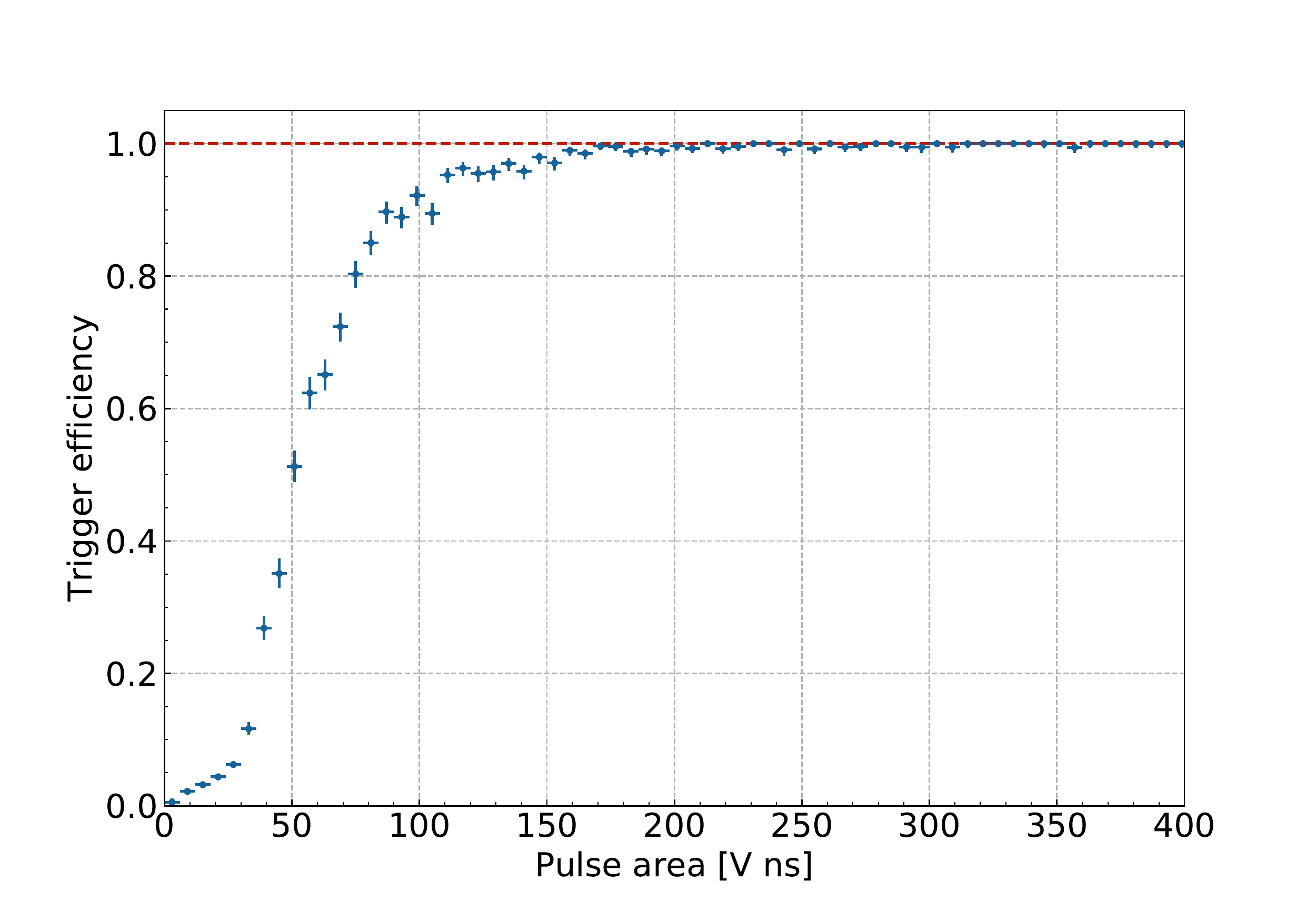}
        \caption{Hardware trigger efficiency as a function of pulse area in the Arktis detector. Experimental trigger efficiency values are applied to the MC spectra when fitting to the Arktis detector data.}
        \label{fig:efficiency}
    \end{figure}
    
    Comparison of MC results to experimental data required an understanding of the trigger efficiency of the Arktis detector as a function of the pulse size observed in the detector. The trigger efficiency for detecting two or more pulses in the Arktis detector, shown in Figure \ref{fig:efficiency}, was measured with a modified experimental geometry and modified random triggering approach. Compared to the procedure described in section \ref{ssec:neutron_scattering}, we placed the Arktis detector closer to the DD generator source plane, and we triggered the digitizer on the source pulse of the DD generator running at 2 kHz with a 5\% duty cycle. We recorded the four Arktis channels and the trigger logic signal arising from two coincident Arktis pulses as in the first two-fold coincidence module of Figure~\ref{fig:DAQ_Cartoon}. The efficiency was determined by the fraction of events of a given pulse size, defined as the total area of the waveforms across all four Arktis channels, coincident with an Arktis trigger logic signal. This method of triggering on the source pulse was a more efficient way to collect waveforms with neutron scattering events than a simple random trigger.
    
    \subsection{Time Resolution Measurement}
    
    We used a $^{22}$Na gamma-ray source to trigger the two detectors in coincidence to measure the time resolution of the setup. While gammas have a reduced probability of triggering the Arktis detector, direct exposure to a gamma source still provided a reasonable event rate. The detectors were positioned with the $^{22}$Na source in the middle since $^{22}$Na decays produce back-to-back 511 keV gammas from positron annihilation; the distance of each detector from the $^{22}$Na source was selected so the solid angle they subtended ensured that any gamma pair which triggered the Arktis detector could also trigger the organic scintillator detector \cite{GP2012a}. Pulse times were calculated by using the time of the first sample greater than 25\% of the pulse maximum, and for the Arktis detector we used the earliest pulse time in the four channels. We took the difference in the nominal pulse times for each of the two detectors and fit a Gaussian to the distribution of differences to find a mean offset and standard deviation. We measured the standard deviation in differences, corresponding to the time resolution for our setup, to be $\sigma_t =34$ ns. Our timing convention assumes the neutron scatters in the Arktis detector first, so we calculate the time of flight as the organic scintillator time minus the Arktis detector time; we found a mean offset of $\mu_t = -16$ ns in our measurement. We applied the resolution and mean offset to the MC events by resampling the time of flight of the simulated events from a normal distribution with the measured mean and standard deviation.

\section{Results}
    
    \subsection{Data Selection}
    
    Two basic data selection cuts were applied to discriminate signal events from backgrounds consisting of non-neutron particles scattering in one or both of the detectors and creating accidental coincidences. Additionally, our hardware trigger for the Arktis detector was close to its baseline noise, increasing the rate of events not caused by neutrons scattering once in each detector.
    
\begin{figure}[h]
        \centering\includegraphics[width=5.in]{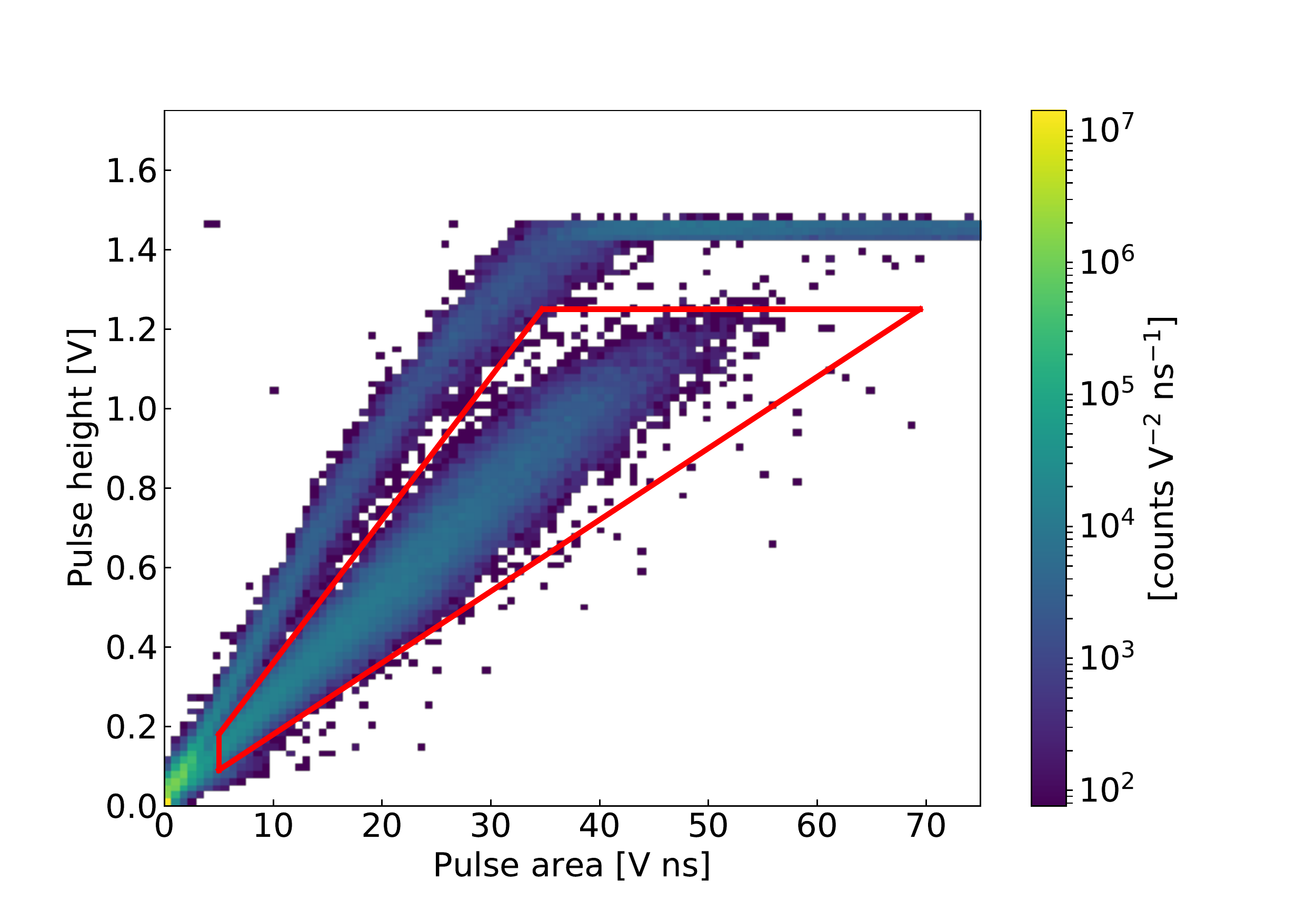}
        \caption{Pulse shape discrimination in the St.~Gobain BC-501A organic scintillator \cite{SG2019a} detector using peak height vs.~pulse area. The lower band corresponds to neutron recoils in the detector; events inside of the red bounding lines were accepted in the analysis. Saturation of the digitizing electronics can be seen at a peak height of about 1.4 V.}
        \label{fig:OS_PSD}
    \end{figure}

    The St.~Gobain BC-501A organic scintillator used to tag scattered neutrons has strong pulse shape discrimination between nuclear recoils and electron recoils. These event types form two distinct bands in pulse height vs.~pulse area, as shown in Figure \ref{fig:OS_PSD}, so we select neutron events by accepting only events in the lower band. There is a minimum area cut below which two bands overlap with each other. This cut eliminates about 90\% of the recorded events, serving as the main method to prevent accidental coincidences from entering the analysis. From Figure \ref{fig:OS_PSD}, it is clear there are a significant number of events triggered by gamma scattering in the organic scintillator detector. These gammas could originate from the DD generator directly, from neutron capture in materials in and around the experiment, or simply from environmental backgrounds. Additionally, some good neutron events are outside of the acceptance region in Figure \ref{fig:OS_PSD} because the recoil energy in the organic scintillator detector was not enough to allow for good pulse shape discrimination.
    
    We also apply a time of flight cut to the data and each corresponding set of MC events, for which the time of flight has been smeared according to the measured time resolution. This cut helps eliminate events that are accidental coincidences and neutrons which scatter multiple times in the Arktis detector and surrounding components. Since the time resolution is comparable to the time of flight of the scattered neutrons, we select events that are +3/$-$1 $\sigma_t$ around the expected time of flight of neutrons at each scattering angle. The asymmetry in the acceptance window was chosen to eliminate trailing multiple scattering neutrons from the dataset while accounting for time resolution.
    
    Combined, the cuts result in an acceptance of 1-5\% of the total recorded events for the five datasets. The results from these cuts are consistent with our expectation of events caused by baseline fluctuations in the Arktis detector, since the smallest acceptance fraction came from lowest-energy recoil dataset, where the neutron event pulses have the lowest trigger efficiency.
    
    \subsection{Fitting Procedure}
    
    After the data selection cuts, we fit the MC spectra to each scattering dataset independently. We assume that the energy resolution is Gaussian and that it scales as the square root of the recoil energy times a constant factor, $A$, which has units $\sqrt{\mathrm{keV}}$, to account for light production and detection efficiency in the detector, as we did not simulate these processes directly. We also take the MC output energy to the digitizer units with a scale factor $C$ with units V ns / keV. It is this factor $C$ that serves as our measurement of the response linearity, since it should have a consistent value across a range of recoil energies given a linearly responding target material.
    
    The MC events were smeared in energy and scaled to the digitizer units by resampling from the the Gaussian probability distribution
    \begin{equation}
        p(x) = \frac{1}{\sqrt{2\pi C^2A^2E}}\exp\left[\frac{-(x-CE)^2}{2C^2A^2E}\right],
    \end{equation}
    where $x$ is the simulated detector response in V ns and $E$ is the energy of the MC event in keV, and we use $\chi^2$ minimization to find the best values of the parameters $A$ and $C$.
    
    The smeared MC events were binned in pulse area, and an experimental measure of the hardware trigger efficiency, shown in Figure \ref{fig:efficiency}, was applied to the spectrum. The resulting MC spectra were compared to corresponding experimental datasets by computing
    \begin{equation}
        \chi^2 = \sum_{i=1}^{N}\frac{(n_i-\nu_i)^2}{\nu_i},
        \label{eq:chi2}
    \end{equation}
    where $n_i$ is the number of experimental events and $\nu_i$ is the number of simulated events in the $i$th bin. The fitting region was first selected by hand to approximate the single-scattering signal region. The MC spectrum in the fitting region was normalized to the experimental spectrum before computing $\chi^2$. This fitting procedure, from MC smearing through computing $\chi^2$, was repeated for a range of energy resolution factors, $A$, and MC scale factors, $C$ in a two dimensional scan of the parameters. The best fit parameters for each dataset were those corresponding to the minimum $\chi^2$ value.
    
\begin{figure}[h!]
        \includegraphics[width=\linewidth]{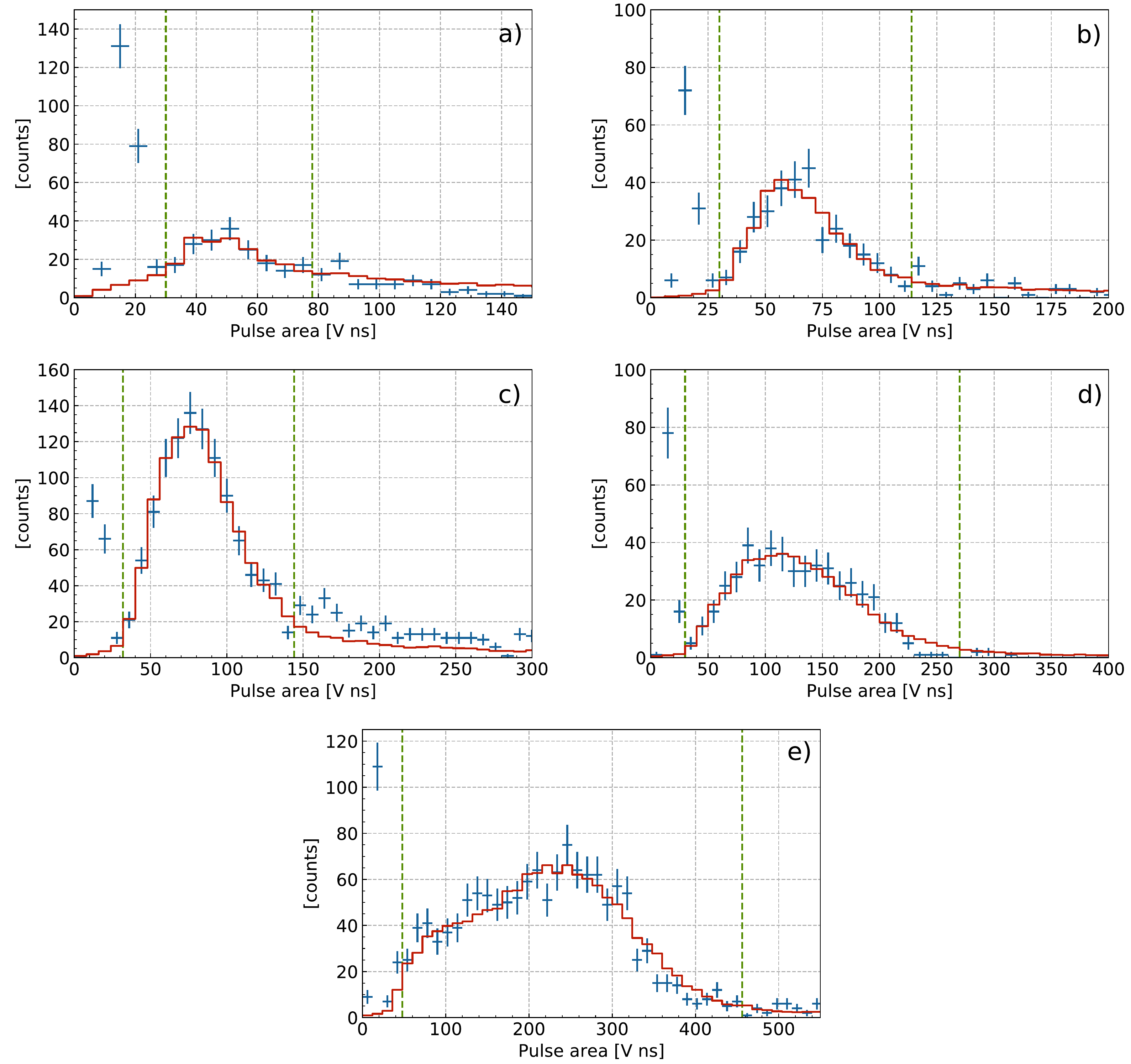}
        \caption{Plots of the experimental data (blue) and fitted Monte Carlo spectra (red) for all five scattering angles. The fit region is between the vertical green dashed lines. The scattering angle (energy) is \textbf{a)} 20$^\mathrm{o}$ (83 keV) \textbf{b)} 25$^\mathrm{o}$ (129 keV) \textbf{c)} 30$^\mathrm{o}$ (182 keV) \textbf{d)} 45$^\mathrm{o}$ (384 keV) and \textbf{e)} 60$^\mathrm{o}$ (626 keV).}
        \label{fig:individual_fits}
    \end{figure}

    After this first pass parameter scan, a smeared MC distribution was produced using the best fit $A$ and $C$. A Gaussian was fit to the distribution of single-scattering neutrons without the application of the trigger efficiency. The fitting region was redefined as $\pm 3\sigma$ of the fit Gaussian around its mean to minimize bias in selecting the fitting region. It was also bounded from below with the same minimum value of 30 V ns for each scattering angle to exclude an extraneous population of low-area events, which can be seen in Figure \ref{fig:individual_fits}. These events are likely noise triggers or electron recoils in the Arktis detector accidentally coincident with neutron events in the organic scintillator detector, since they are fairly consistent in number throughout the datasets and the total population of these events is much larger before the time of flight cut. For the new fit region, MC smearing and computing $\chi^2$ was repeated for the same range of parameters, and the reported best fit parameters $A$ and $C$ were those found after this second pass. Experimental data and their best fits are shown in Figure \ref{fig:individual_fits}.
    
    \subsection{Error Analysis}
    
    There were several effects considered for estimating the errors in the reported best fit values. Statistical error was obtained directly from the $\chi^2$ fits. To estimate the systematic errors, we evaluated the effect on the best fit parameters when changing the pulse timing parameters, data selection cuts, and MC geometry in the analysis procedure above. For each systematic error fit, the energy resolution factor was capped at $A = 13~\sqrt{\mathrm{keV}}$, roughly twice the average energy resolution factor found in the main analysis fits. This restriction was imposed so that the energy resolution factor $A$ did not diverge too far from the values found at high recoil energy, which yielded more consistent results. The systematic error associated with each variation was taken as the difference between the best fit value in the original analysis and after the modification was applied. 
    
    The timing of smaller pulses in the Arktis detector could be strongly affected by the fractional value of the maximum pulse height chosen to determine the pulse time, while the offset associated with the rise time of larger pulses could also be affected by this value. We used 10\% and 50\% of the maximum value as alternate values for the pulse timing threshold. We also used the second pulse time of the Arktis detector instead of the first to evaluate the systematic in calculating the neutron time of flight to consider whether there was a systematic effect from stray pulses. The organic scintillator detector pulse shape discrimination minimum area cut was modified by $\pm 50\%$ to study the impact of the band separation. The accepted time of flight region was expanded to +4/$-$4 $\sigma_t$ and contracted to +2/$-$0 $\sigma_t$. To better understand the effect of the fit region, the number of deviations of the single-scatter peak that defined the fit region were decreased to $\pm 2\sigma$ and increased to $\pm 4\sigma$. 
    
    We determined an uncertainty of 0.5 cm in the measured positions of the two detectors to evaluate systematics associated with uncertainties in the detector positions used in the MC geometry. We propagated these errors into errors in the neutron scattering angle and distance between the detectors and generated MC events according to the modified angle and distance for a total of four additional MC datasets for each scattering angle.
    
\begin{figure}[h]
        \centering\includegraphics[width=5.in]{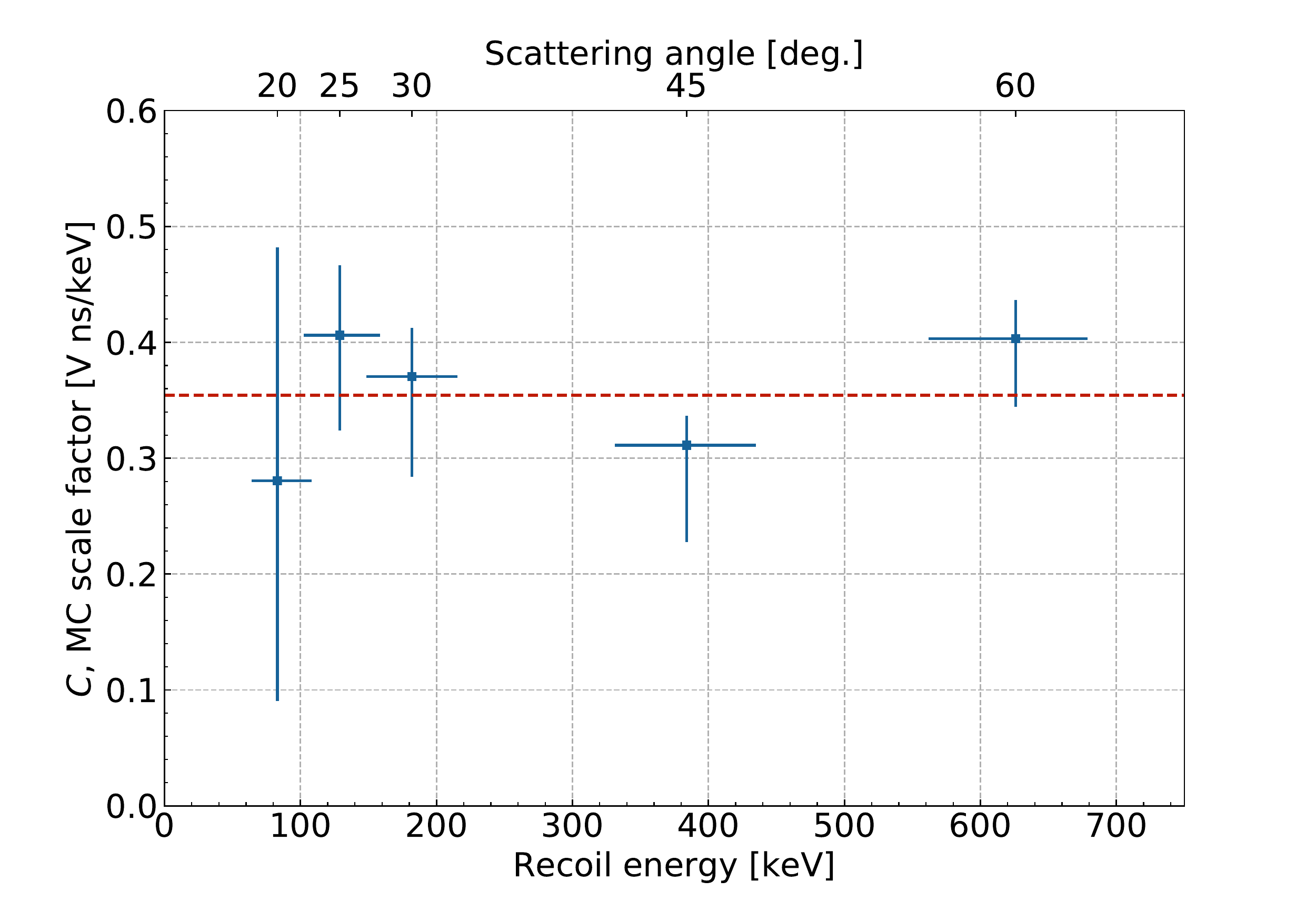}
        \caption{Best fit MC scale factor, $C,$ for each scattering angle, shown in terms of the corresponding neutron recoil energy in the Arktis detector calculated in Equation~\ref{eq:energy_angle}. Best fit scale factors are shown in blue, and the mean value of all the data points is the dashed red line. The horizontal error bars correspond to the spread in angles/energies given in Table \ref{tab:energy_ranges}. The vertical error bars are from the combination of statistical and systematic uncertainties and are reported in Table \ref{tab:errors}. }
        \label{fig:linearity}
    \end{figure}

    The individual systematic errors were ultimately combined in quadrature for a total systematic error estimate for each scattering angle. Then, the systematic and statistical errors were combined in quadrature for an estimate of the total error in the best fit parameters. The combined total errors from this analysis, which are uniformly dominated by the systematic errors, can be seen in Figure \ref{fig:linearity} and Table \ref{tab:errors}.
        
    \begin{table}[h]
        \centering
        \caption{Best fit values of the energy resolution factor, $A$, and MC scale factor, $C$, and their errors for each scattering angle $\theta$ corresponding to a recoil energy $E_r$. Errors are quadratic sum of the $1\sigma$ statistical errors from the $\chi^2$ fit and systematic errors found as reported in the text. The $\chi^2$ results and number of degrees of freedom for each fit are also reported.}
        \begin{tabular}{|c
                        c
                        c
                        c
                        c|}
            \hline
            $\theta$ & $E_r$ & $A$ & $C$ & $\chi^2$ / $\mathrm{DoF}$\rule{0pt}{2.6ex}\\
            $[\mathrm{deg.}]$ & $[\mathrm{keV}]$ & $[\sqrt{\mathrm{keV}}]$ & $[\mathrm{V\ ns/keV}]$ &\\
            \hline\hline\rule{0pt}{2.6ex}
            20 &  83 & $6.4_{-6.0}^{+9.7}$ & $0.28_{-0.19}^{+0.20}$ & 2.044 / 5\\[3pt]
            25 & 129 & $3.6_{-0.8}^{+1.3}$ & $0.41_{-0.08}^{+0.06}$ & 10.18 / 11\\[3pt]
            30 & 182 & $5.4_{-1.1}^{+4.5}$ & $0.37_{-0.09}^{+0.04}$ & 10.71 / 11\\[3pt]
            45 & 384 & $8.4_{-1.7}^{+5.8}$ & $0.31_{-0.08}^{+0.03}$ & 22.89 / 21\\[3pt]
            60 & 626 & $6.9_{-2.2}^{+3.8}$ & $0.40_{-0.06}^{+0.03}$ & 49.12 / 32\\[3pt]
            \hline
        \end{tabular}
        \label{tab:errors}
    \end{table}

    \subsection{Discussion}

    The best fit MC scale factors have been plotted in Figure~\ref{fig:linearity}, along with a horizontal line showing their mean. If the scintillation response of the detector is linear, the same MC scale factor, $C,$ should provide the best fit across the range of recoil energies. Besides the scintillation behavior of the helium gas itself, nonlinearities in the response could be introduced by the onboard readout electronics of the Arktis detector and possible non-homogeneity in the light collection efficiency of the detector. While we attempted to model the light production physics and the light collection efficiency of the detector using simple Gaussian smearing proportional to the square root of the recoil energy, it is possible that a more detailed simulation of the optical processes in the detector could yield better fits to the data. Still, each scattering angle was fit to a reasonable $\chi^2$ per degree of freedom and the best fit values of $C$ deviated at most 21\% from the mean of 0.35 V ns/keV.
    
    At the lowest energies, we do not resolve the fitted parameters well because the effects of the energy resolution and scaling factor compete with the effects of the threshold efficiency. This analysis does not provide much traction on the energy resolution factor, which is large most likely due to a combination of the light collection efficiency of the detector and the onboard shaping electronics used for signal amplification. A common alternative analysis used with well-characterized neutron sources involves a precise measurement of the neutron time of flight to determine the recoil energy in the detector. However, our measured time resolution of $\sigma_t =34$ ns is much too large to allow for a determination of the energy in this way. For geometries comparable to those used in this experiment, single-nanosecond time resolution would provide similar energy resolution to our measured values. Such time resolution would be difficult to achieve given the 4 ns sample rate of our digitizing electronics, making the geometrical determination of the neutron energy advantageous for the neutron energies and distances travelled in this experiment. 
    
    We conclude by noting that 83 keV is about the energy threshold for neutron recoils that can be probed by the detector. An $^{88}$Y/Be photoneutron source produces 153 keV neutrons, which according to Equation \ref{eq:energy_angle} can deposit at most 97 keV in a single elastic scattering interaction with a helium atom. This version of the Arktis S-670 detector could in principle detect the high end of neutron recoils produced by an $^{88}$Y/Be source and, in combination with detailed simulations of the source and detector geometry, provide sufficient sensitivity to characterize the source activity.
        
\section{Conclusion}

    We measured the scintillation response of a high pressure $^4$He gas scintillation detector to nuclear recoils using monoenergetic neutrons from a DD generator. We used a second detector, an organic liquid scintillator volume instrumented with a PMT, to tag neutrons scattering into a series of angles, thereby fixing the recoil energy deposited into the helium gas. We extracted the linearity of the scintillation response to these energy deposits by fitting Monte Carlo spectra to the measured data and finding the scale factor resulting in the best fit. We found that the detector is fairly linear in its response across a range of recoil energies between 83-626 keV with the best fit values of the scale factor. This result suggests the detector might be applied towards characterizing low-energy neutron sources by taking advantage of its low energy threshold and the response linearity of helium gas. Further work in characterizing the optical response of the detector and pulse integration could help improve the detector's performance in detecting low-energy neutrons and determining their energy spectra.
    
\acknowledgments

    This work was supported in part by DOE grants DE-SC0018982, DE-SC0019319, and Quantum Information Science Enabled Discovery (QuantISED) for High Energy Physics (KA2401032). V. Velan is supported by a DOE Graduate Instrumentation Research Award. This material is based upon work supported by the National Science Foundation Graduate Research Fellowship under Grant No. DGE 1106400. We thank G.~Davatz and U.~Gendotti for helpful discussions about the Arktis S-670 detector. This research used the Savio computational cluster resource provided by the Berkeley Research Computing program at the University of California, Berkeley (supported by the UC Berkeley Chancellor, Vice Chancellor for Research, and Chief Information Officer).

\bibliographystyle{JHEP}
\bibliography{arktis.bib}

\end{document}